\documentclass[twocolumn,showpacs,superscriptaddress,amsmath,amssymb]{revtex4}
\usepackage{graphicx}
\usepackage{dcolumn}
\usepackage{bm}

\def\bd{
\begin{document}} \def\ed{\end{document}}
\def\bmp{\begin{minipage}} \def\emp{\end{minipage}}
\def\bcc{\begin{center}} \def\ecc{\end{center}}     \def\npg{\newpage}
\def\beq{\begin{equation}} \def\eeq{\end{equation}} \def\hph{\hphantom}
\def\be{\begin{equation}} \def\ee{\end{equation}} \def\r#1{$^{[#1]}$}
\def\n{\noindent} \def\ni{\noindent} \def\pa{\parindent}
\def\hs{\hskip} \def\vs{\vskip} \def\hf{\hfill} \def\ej{\vfill\eject}
\def\cl{\centerline} \def\ob{\obeylines}  \def\ls{\leftskip}
\def\underbar#1{$\setbox0=\hbox{#1} \dp0=1.5pt \mathsurround=0pt
   \underline{\box0}$}   \def\ub{\underbar}    \def\ul{\underline}
\def\f{\left} \def\g{\right} \def\e{{\rm e}} \def\o{\over} \def\d{{\rm d}}
\def\vf{\varphi} \def\pl{\partial} \def\cov{{\rm cov}} \def\ch{{\rm ch}}
\def\la{\langle} \def\ra{\rangle} \def\EE{e$^+$e$^-$} \def\pt{p_{\rm t}}
\def\bitz{\begin{itemize}} \def\eitz{\end{itemize}}
\def\btbl{\begin{tabular}} \def\etbl{\end{tabular}}
\def\btbb{\begin{tabbing}} \def\etbb{\end{tabbing}}
\def\beqar{\begin{eqnarray}} \def\eeqar{\end{eqnarray}}
\def\\{\hfill\break} \def\dit{\item{-}} \def\i{\item}
\def\bbb{\begin{thebibliography}{9} \itemsep=-1mm}
\def\ebb{\end{thebibliography}} \def\bb{\bibitem}
\def\bpic{\begin{picture}(260,240)} \def\epic{\end{picture}}
\def\akgt{\cl{\bf ACKNOWLEDGMENTS}}
\def\fgn{\noindent{\bf\large\bf figure captions}}
\def\lan{\langle}
\def\ran{\rangle}
\def\p{\pi}
\def\ifmath#1{\relax\ifmmode #1\else $#1$\fi}%
\def\rc{\ifmath{{\mathrm{c}}}}
\def\cut{\ifmath{{\mathrm{cut}}}}
\def\rF{\ifmath{{\mathrm{F}}}}
\def\rK{\ifmath{{\mathrm{K}}}}
\def\rp{\ifmath{{\mathrm{p}}}}
\def\rt{\ifmath{{\mathrm{t}}}}
\def\LAB{\ifmath{{\mathrm{LAB}}}}
\def\cut{\ifmath{{\mathrm{cut}}}}
\def\beq{\begin{equation}}
\def\eeq{\end{equation}}

\newcommand{\cinst}[2]{$^{\mathrm{#1}}$~#2\par}
\newcommand{\crefi}[1]{$^{\mathrm{#1}}$}
\newcommand{\crefii}[2]{$^{\mathrm{#1,#2}}$}
\newcommand{\crefiii}[3]{$^{\mathrm{#1,#2,#3}}$}
\newcommand{\HRule}{\rule{0.5\linewidth}{0.5mm}}

\bd

\title{\boldmath Confirmation of the $X(1835)$ and observation of the resonances $X(2120)$ and $X(2370)$
in $J/\psi\rightarrow\gamma\pi^+\pi^-\eta^\prime$}


\author{\small
M.~Ablikim$^{1}$, M.~N.~Achasov$^{5}$, L.~An$^{9}$, Q.~An$^{36}$,
Z.~H.~An$^{1}$, J.~Z.~Bai$^{1}$, R.~Baldini$^{17}$, Y.~Ban$^{23}$,
J.~Becker$^{2}$, N.~Berger$^{1}$
% ...
\\
\vspace{0.2cm}
(BESIII Collaboration)\\
\vspace{0.2cm} {\it
$^{1}$ Institute of High Energy Physics, Beijing 100049, P. R. China\\
$^{2}$ Bochum Ruhr-University, 44780 Bochum, Germany\\
% ...
$^{5}$ G.I. Budker Institute of Nuclear Physics SB RAS (BINP), Novosibirsk 630090, Russia\\
% ...
$^{9}$ Guangxi University, Naning 530004, P. R. China\\
% ...
$^{17}$ Laboratori Nazionali di Frascati - INFN, 00044 Frascati, Italy\\
% ...
$^{23}$ Peking University, Beijing 100871, P. R. China\\
% ...
$^{36}$ University of Science and Technology of China, Hefei 230026, P. R. China\\
% ...
}}

\vspace{0.4cm}

\date{\today}

\begin{abstract}

 With a sample of $(225.2\pm2.8)\times10^{6}$ $J/\psi$ events registered in the
BESIII detector, $J/\psi\rightarrow\gamma\pi^{+}\pi^{-}\eta^\prime$
is studied using two $\eta^{\prime}$ decay modes:
$\eta^\prime\rightarrow\pi^{+}\pi^{-}\eta$ and
$\eta^\prime\rightarrow\gamma\rho^{0}$. The $X(1835)$, which was
previously observed by  BESII, is confirmed with a statistical
significance that is larger than $20\sigma$. In addition, in the
$\pi^+\pi^-\eta^\prime$ invariant mass spectrum, the $X(2120)$ and
the $X(2370)$, are observed with statistical significances larger
than ~$7.2\sigma$~ and ~$6.4\sigma$~, respectively. For the
$X(1835)$, the angular distribution of the radiative photon is
consistent with expectations for a pseudoscalar.

\end{abstract}


\maketitle

A $\pi^+\pi^-\eta^\prime$ resonance, the $X(1835)$, was observed in
$J/\psi\rightarrow\gamma\pi^{+}\pi^{-}\eta^\prime$ decays with a
statistical significance of $7.7\sigma$ by the BESII
experiment~\cite{x1835}. A fit to a Breit-Wigner function yielded a
mass $M=1833.7\pm6.1({\rm stat})\pm2.7({\rm syst})~{\rm MeV}/c^{2}$,
a width $\Gamma=67.7\pm20.3({\rm stat})\pm7.7({\rm syst})~{\rm
MeV}/c^{2}$, and a product branching fraction
$B(J/\psi\rightarrow\gamma X)\cdot
B(X\rightarrow\pi^{+}\pi^{-}\eta\prime)=(2.2\pm0.4({\rm
stat})\pm0.4({\rm syst}))\times10^{-4}$.  The study was stimulated
by the anomalous  $p\bar{p}$  invariance mass threshold enhancement,
that was reported in $J/\psi\rightarrow\gamma p\bar{p}$ decays by
the BESII experiment~\cite{ppb_jixb} and was recently confirmed in
an analysis of
$\psi^\prime\rightarrow\pi^+\pi^-J/\psi, ~J/\psi\rightarrow\gamma
p\bar{p}$ decays by the BESIII experiment~\cite{ppb_bes3}. The
possible interpretations of the $X(1835)$ include a $p\bar{p}$ bound
state~\cite{theory1,theory2,theory4,theory5}, a glueball
~\cite{theory6,theory7,theory8}, a radial excitation of the
$\eta^\prime$ meson~\cite{theory9}, etc. A high statistics data
sample collected with BESIII provides an opportunity to confirm the
existence of the $X(1835)$  and look for possible related states
that decay to $\pi^+\pi^-\eta^{\prime}$.

Lattice QCD predicts that the lowest lying pseudo-scalar glueball
meson has a mass that is around
 ~$2.3~{\rm GeV}/c^2$ ~\cite{LQCD}. This pseudo-scalar
glueball may have properties in common with the ~$\eta_c$~, due to
its similar decay dynamics that favor decays into gluons. One of the
strongest decay channels of the ~$\eta_c$~ is
~$\pi^+\pi^-\eta^\prime$~. Thus
~$J/\psi\rightarrow\gamma\pi^+\pi^-\eta^\prime$~ decays may be a
good channel for finding ~$0^{-+}$~ glueballs.

In this letter, we report a study of
$J/\psi\rightarrow\gamma\pi^+\pi^-\eta^\prime$ that uses two
$\eta^\prime$ decay modes,
 $\eta^\prime\rightarrow\gamma\rho$ and $\eta^\prime\rightarrow\pi^+\pi^-\eta$. The analysis
uses a sample of $(225.2\pm2.8)\times10^{6}$  $J/\psi$ events
~\cite{jpsi} accumulated in the new Beijing Spectrometer
(BESIII)~\cite{bes3} located at the Beijing Electron-Positron
Collider (BEPCII)~\cite{bes2} at the Beijing Institute of High
Energy Physics.

The design peak luminosity of the BEPCII double-ring $e^+e^-$
collider, is $10^{33}$ cm$^{-2}s^{-1}$ with beam currents of 0.93~A.
The BESIII detector has a geometrical acceptance of 93\% of 4$\pi$
and consists of four main components. 1) A small-celled,
helium-based main draft chamber (MDC) with 43 layers.
The average single wire resolution is 135 $\mu$m, and the momentum
resolution for 1~GeV$/c$ charged particles in a 1 T magnetic field
is 0.5\%. 2) An electromagnetic calorimeter (EMC) comprised of 6240
CsI (Tl) crystals arranged in a cylindrical shape (barrel) plus two
endcaps.  The energy resolution for 1.0~GeV photons is 2.5\% in the
barrel and 5\% in the endcaps, and the position resolution is 6~mm
in the barrel and 9~mm in the endcaps. 3) A Time-Of-Flight system
(TOF) for particle identification composed of a barrel part with
two layers with 88 pieces of 5~cm thick, 2.4~m long plastic
scintillators in each layer, and two endcaps each with 96 fan-shaped,
5~cm thick, plastic scintillators.  The time
resolution is 80~ps in the barrel, and 110 ps in the endcaps,
corresponding to a $2\sigma$ K/$\pi$ separation for momenta up to
1.0~GeV$/c$; 4) A muon chamber system (MUC) made of 1000~m$^2$ of
Resistive Plate Chambers (RPC) arranged in 9 layers in the barrel
and 8 layers in the endcaps and incorporated in the return iron of
the superconducting magnet. The position resolution is about 2~cm.

Charged-particle tracks in the polar angle range $|\cos\theta|<0.93$
are reconstructed from hits in the MDC. Tracks that extrapolate to
 be within $20~{\rm cm}$ of the
interaction point in the beam direction and  $2~{\rm cm}$ in the
plane perpendicular to the beam are selected.  The TOF and $dE/dx$
information are combined to form particle identification confidence
levels for the $\pi$, $K$, and $p$ hypotheses; each track is
assigned to the particle type that corresponds to the hypothesis
with the highest confidence level.
 Photon candidates are required to have at least $100~{\rm MeV}$ of energy in the
EMC regions $|\cos\theta|<0.8$ and $0.86<|\cos\theta|<0.92$ and be
isolated from all charged tracks by more than $5^\circ$. In this
analysis, candidate events are required to have four charged tracks
(zero net charge) with at least three of the charged tracks
identified as pions. At least two photons (three photons)  are
required for the $\eta^\prime\rightarrow\gamma\rho$
($\eta^\prime\rightarrow\pi^+\pi^-\eta$) channel.

For
 $J/\psi\rightarrow\gamma\pi^+\pi^-\eta^\prime(\eta^\prime\rightarrow\gamma\rho)$,
 a four-constraint (4C) energy-momentum conservation kinematic fit is
 performed to the $\gamma\gamma\pi^+\pi^-\pi^+\pi^-$ hypothesis.  For
 events with more than two photon candidates, the combination with the
 minimum $\chi^2$ is used, and $\chi^2_{4C}<40$ is required.  Events
 with $|M_{\gamma\gamma}-m_{\pi^0}|<0.04~{\rm GeV}/c^2$,
 $|M_{\gamma\gamma}-m_{\eta}|<0.03~{\rm GeV}/c^2$, $0.72~{\rm
 GeV}/c^2<M_{\gamma\gamma}<0.82~{\rm GeV}/c^2$ or
 $|M_{\gamma\pi^+\pi^-}-m_{\eta}|<0.007{~\rm GeV}/c^2$ are rejected to
 suppress the background from $\pi^0\pi^+\pi^-\pi^+\pi^-$,
 $\eta\pi^+\pi^-\pi^+\pi^-$,
 $\omega(\omega\rightarrow\gamma\pi^0)\pi^+\pi^-\pi^+\pi^-$ and
 $\gamma\pi^+\pi^-\eta(\eta\rightarrow\gamma\pi^+\pi^-)$, respectively.
 A clear $\eta^\prime$ signal with a $5~{\rm MeV}/c^2$ mass resolution
 is evident in the mass spectrum of all selected $\gamma\pi^+\pi^-$
 combinations shown in Fig.~\ref{Fig1}(a). Candidate $\rho$ and
 $\eta^\prime$ mesons are reconstructed from the $\pi^+\pi^-$ and
 $\gamma\pi^+\pi^-$ pairs with $|M_{\pi^+\pi^-}-m_\rho|<0.2~{\rm
 GeV}/c^2$ and $|M_{\gamma\pi^+\pi^-}-m_{\eta^\prime}|<0.015{~\rm
 GeV}/c^2$, respectively. If more than one combination passes these
 criteria, the combination with $M_{\gamma\pi^+\pi^-}$ closest to
 $m_{\eta^\prime}$ is selected. After the above selection, the
 $X(1835)$ resonance is clearly visible in the $\pi^+\pi^-\eta^\prime$
 invariant mass spectrum of Fig.~\ref{Fig1}(b). Also, additional peaks
 are evident around 2.1 and 2.4$~{\rm GeV}/c^2$ as well as a distinct
 signal for the $\eta_c$.

\begin{figure}
\includegraphics[width=3.5in]{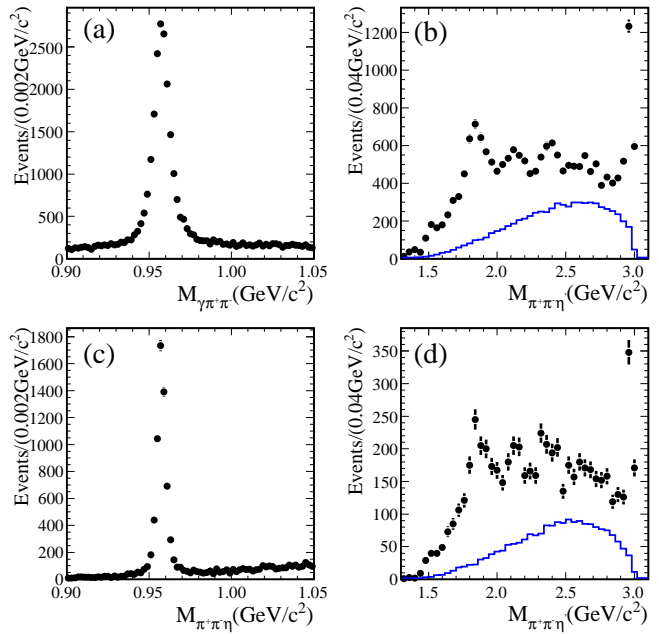}
\caption{\label{Fig1} Invariant-mass distributions for the selected
candidate events.  $(a)$ and $(b)$ are the $\gamma\pi^+\pi^-$
invariant-mass spectrum and the $\pi^+\pi^-\eta^\prime$ invariant-mass
spectrum for $\eta^\prime\rightarrow\gamma\rho$, respectively.  $(c)$
and $(d)$ are the $\pi^+\pi^-\eta$ invariant-mass spectrum and the
$\pi^+\pi^-\eta^\prime$ invariant-mass spectrum for
$\eta^\prime\rightarrow\pi^+\pi^-\eta$, respectively. The histograms
in (b) and (d) are from $J/\psi\rightarrow\gamma\pi^+\pi^-\eta^\prime$
phase-space MC events (with arbitrary normalization) for
$\eta^\prime\rightarrow\gamma\rho$ and
$\eta^\prime\rightarrow\pi^+\pi^-\eta$, respectively.}
\end{figure}

For
$J/\psi\rightarrow\gamma\pi^+\pi^-\eta^\prime(\eta^\prime\rightarrow\pi^+\pi^-\eta)$,
a 4C kinematic fit to the $\gamma\gamma\gamma\pi^+\pi^-\pi^+\pi^-$
hypothesis is performed. If there are more than three photon
candidates, the combination with
the minimum $\chi^2_{4C}$ is selected, and $\chi^2_{4C}<40$ is required. In
order to reduce the combinatorial background events from
$\pi^0\rightarrow\gamma\gamma$,
$|M_{\gamma\gamma}-m_{\pi^0}|>0.04{~\rm GeV}/c^2$ is required for all
photon pairs. The $\eta$ candidates are selected by requiring
$|M_{\gamma\gamma}-m_\eta|<0.03{~\rm GeV}/c^2$. A five-constraint (5C)
fit with an $\eta$ mass constraint is used to improve the mass
resolution from $8~{\rm MeV}/c^2$(4C) to $3~{\rm MeV}/c^2$, as shown in
Fig.~\ref{Fig1}(c) where $\chi^2_{5C}<40$ is required. To select
$\eta^\prime$ mesons,
$|M_{\pi^+\pi^-\eta}-m_{\eta^\prime}|<0.01{~\rm GeV}/c^2$ is
required. If more than one combination passes the above selection, the
combination with $M_{\pi^+\pi^-\eta}$ closest to $m_{\eta^\prime}$ is
selected. After the above selection, structures similar to those seen
for the $\eta^\prime\rightarrow\gamma\rho$ channel in the
$\pi^+\pi^-\eta^\prime$ invariant mass spectrum can be seen in
Fig.~\ref{Fig1}(d), namely peaks near 1.8, 2.1 and 2.4 ${~\rm
GeV}/c^2$ as well as the $\eta_c$.

Potential background processes are studied with an inclusive sample of
$2\times10^8$ $J/\psi$ events generated according to the Lund-Charm
model~\cite{lund} and the Particle Data Group (PDG) decay
tables~\cite{PDG2}. There are no peaking backgrounds at the positions
of the three resonances. To ensure further that the three peaks are
not due to background, we have studied potential exclusive background
processes using data.  The main background channel is from
$J/\psi\rightarrow\pi^0\pi^+\pi^-\eta^\prime$. Non-$\eta^\prime$
processes are studied with $\eta^\prime$ mass-sideband events. Neither
of these produce peaking structures.

The $\pi^+\pi^-\eta^\prime$ invariant mass spectrum for the combined
two $\eta^\prime$ decay modes is presented in Fig.~\ref{Fig2}. Here
a small peak at the position of the  $f_1(1510)$ signal is also
present. Fits to the mass spectra have been made using four
efficiency-corrected Breit-Wigner functions convolved with a
Gaussian mass resolution plus a non-resonant
 $\pi^+\pi^-\eta^\prime$ contribution and  background representations,
 where the efficiency for the combined
channels is obtained from the branching-ratio-weighted average of
the efficiencies for the two $\eta^\prime$ modes. The contribution
from non-resonant $\gamma\pi^+\pi^-\eta^\prime$ production is
 described  by reconstructed MC-generated
$J/\psi\rightarrow\gamma\pi^+\pi^-\eta^\prime$ Phase Space (PS)
decays, and it is treated as an incoherent process. The background
contribution can be divided into two different components, the
contribution from non-$\eta^\prime$ events estimated from
$\eta^\prime$ mass sideband, and the contribution from
$J/\psi\rightarrow\pi^0\pi^+\pi^-\eta^\prime$. For the second
background, we obtain the background $\pi^+\pi^-\eta^\prime$ mass
spectrum from data by selecting
$J/\psi\rightarrow\pi^0\pi^+\pi^-\eta^\prime$ events and reweighting their
mass spectrum with  a
weight equal to the MC efficiency ratio of the
$\gamma\pi^+\pi^-\eta^\prime$ and $\pi^0\pi^+\pi^-\eta^\prime$
selections for $J/\psi\rightarrow\pi^0\pi^+\pi^-\eta^\prime$. The
masses, widths and number of event of the $f_1(1510)$, the $X(1835)$
and the resonances near 2.1 and 2.4 ${\rm GeV}/c^2$, the $X(2120)$
and $X(2370)$, are listed in Table \ref{table1}. 
 The statistical significance is
determined from the change in $-2{\rm ln}L$ in the fits to mass
spectra with and without signal assumption while considering the
change of degree of freedom of the fits. With the systematic
uncertainties in the fit taken into account, the statistical
significance of the $X(1835)$ is larger than $20\sigma$, while those
for the $f_1(1510)$, the $X(2120)$ and the $X(2370)$ are larger than
$5.7\sigma$, $7.2\sigma$ and $6.4\sigma$, respectively. The mass and
width from the fit of the $f_1(1510)$ are consistent with PDG
values~\cite{PDG2}. With MC-determined selection efficiencies of
$16.0\%$ and $11.3\%$ for the $\eta^\prime\rightarrow\gamma\rho$ and
$\eta^\prime\rightarrow\pi^+\pi^-\eta$ decay modes respectively, the
branching fraction for the $X(1835)$ is measured to be
 $B(J/\psi\rightarrow\gamma X(1835))\cdot B(X(1835)\rightarrow\pi^{+}\pi^{-}\eta^\prime) = (2.87\pm0.09)\times10^{-4}$.
The consistency between the two $\eta^\prime$ decay modes is checked by fitting their $\pi^+\pi^-\eta^\prime$ mass
distribution separately with the procedure described above.

\begin{figure}
\includegraphics[width=2.8in]{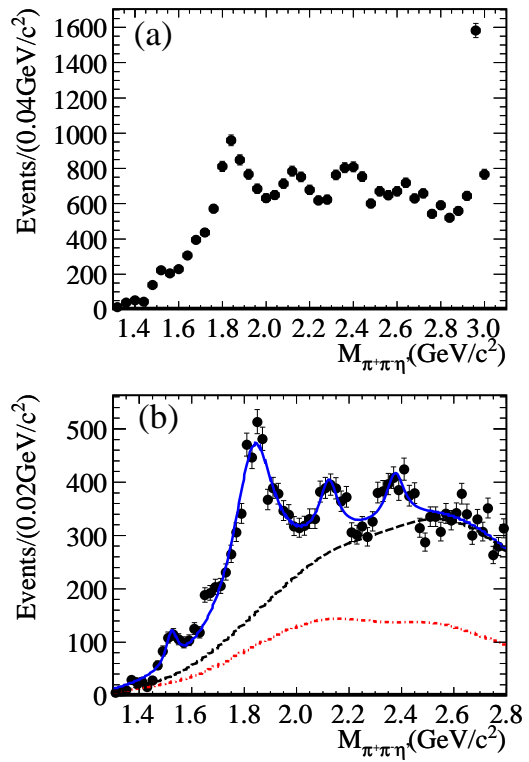}
\caption{\label{Fig2} (a) The $\pi^+\pi^-\eta^\prime$ invariant-mass
distribution for the selected events from the two $\eta^\prime$
decay modes. $(b)$ mass spectrum fitting with four resonances, here,
the dash-dot line is contributions of non-$\eta^\prime$ events and
the $\pi^0\pi^+\pi^-\eta^\prime$ background for two  $\eta^\prime$
decay modes and the dash line is  contributions of the total
background  and non-resonant $\pi^+\pi^-\eta^\prime$ process.}
\end{figure}

\begin{table}[htpb]
\small \caption{Fit results with four resonances for the combined
two $\eta^\prime$ decay modes}
\begin{center}
\begin{tabular}{c|c|c|c}
\hline \hline
resonance &  $M(~{\rm MeV}/c^{2})$ &$\Gamma(~{\rm MeV}/c^{2})$ &$N_{event}$ \\
\hline
$f_1(1510)$ & $1522.7\pm5.0$ & $48\pm11$ & $230\pm37$  \\
\hline
$X(1835)$ & $1836.5\pm3.0$ & $190.1\pm9.0$ & $4265\pm131$  \\
\hline
$X(2120)$ & $2122.4\pm6.7$ & $83\pm16$ & $647\pm103$   \\
\hline
$X(2370)$ & $2376.3\pm8.7$ & $83\pm17$ & $565\pm105$   \\
\hline \hline
\end{tabular}
\end{center}
\label{table1}
\end{table}

For radiative $J/\psi$ decays to a pseudoscalar meson,
    the polar angle of the photon in the $J/\psi$ center of mass system, $\theta_\gamma$,
should be distributed according to $1+\cos^2\theta_\gamma$. We
divide the $|\cos\theta_\gamma|$ distribution into 10 bins in the
region of $[0, 1.0]$. With the same procedure as described above,
the number of the $X(1835)$ events in each bin can be obtained by
fitting the mass spectrum in this bin, and then the
background-subtracted, acceptance-corrected $|\cos\theta_\gamma|$
distribution for the $X(1835)$ is obtained as shown in
Fig.~\ref{Fig3}, where the errors are statistical only. It agrees
with $1+\cos^2\theta_\gamma$, which is expected for a pseudoscalar,
with $\chi^2/d.o.f=11.8/9$.

\begin{figure}
\includegraphics[height=1.8in,width=2.5in]{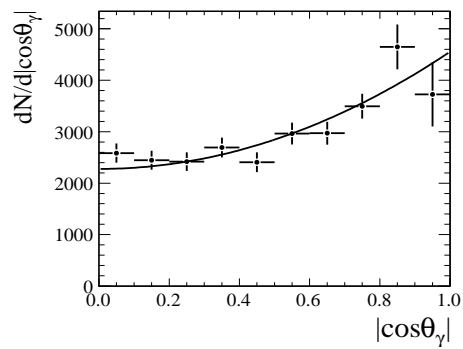}
\caption{\label{Fig3} The background-subtracted,
acceptance-corrected $|\cos\theta_\gamma|$ distribution of the
$X(1835)$ for two $\eta^\prime$ decay modes for
$J/\psi\rightarrow\gamma\pi^+\pi^-\eta^\prime$.}
\end{figure}

The systematic uncertainties on the mass and width are mainly from
the uncertainty of background representation, the mass range
included in the fit, different shapes for background contributions
and the non-resonant process
 and contributions of possible additional resonances in the $1.6{~\rm GeV}/c^2$ and $2.6{~\rm GeV}/c^2$ mass regions. From the study of
$J/\psi\rightarrow p\bar{p}\pi^+\pi^-$, the PID efficiency
difference between data and MC is determined. Using this difference
and reweighting each MC event with a weight equal to the
efficiency ratio between data and MC, we re-fit the mass spectra
and take the changes as systematic uncertainties associated with
data and MC inconsistencies for PID efficiencies. The total
systematic errors on the mass and width are
 $ ^{+5.6}_{-2.1}$ and $ ^{+38}_{-36}$ ${\rm MeV}/c^2$ for the $X(1835)$,
 $ ^{+4.7}_{-2.7}$ and $ ^{+31}_{-11}$ ${\rm MeV}/c^2$ for the $X(2120)$,
 $ ^{+3.2}_{-4.3}$ and $ ^{+44}_{-6}$  ${\rm MeV}/c^2$ for the $X(2370)$ respectively.
For the systematic error of the branching fraction measurement, we
additionally include the uncertainties of the MC generator, charged
track detection efficiency, photon detection efficiency, kinematic
fit, the $\eta^\prime$ decay branching fractions to $\pi^+\pi^-\eta$
and $\gamma\rho$~\cite{PDG2}, the requirement on the $\gamma\gamma$
invariant-mass distribution, signals selection of $\rho$, $\eta$ and
$\eta^\prime$ and the total number of $J/\psi$ events~\cite{jpsi}.
The main contribution also comes from the uncertainty in the
background estimation, and the total relative systematic error on
the product branching fraction for the $X(1835)$ is $
^{+17\%}_{-18\%}$.

In summary, the decay channel $J/\psi\rightarrow\pi^+\pi^-\eta^\prime$ is analyzed using two $\eta^\prime$
decay modes, $\eta^\prime\rightarrow\gamma\rho$ and $\eta^\prime\rightarrow\pi^+\pi^-\eta$. The $X(1835)$,
which was first observed at BESII,
has been confirmed with a statistical significance larger than $20\sigma$.
 Meanwhile, two resonances, the $X(2120)$ and
the $X(2370)$ are observed with statistical significances larger
than $7.2\sigma$ and $6.4\sigma$ respectively. The masses and widths
are measured to be:
 \begin{itemize}
\item $X(1835)$£º

\ \ \  ~$M=1836.5\pm3.0({\rm stat})^{+5.6}_{-2.1}({\rm syst})~{\rm
MeV}/c^2$~

\ \ \  ~$\Gamma=190\pm9({\rm stat})^{+38}_{-36}({\rm syst})~{\rm
MeV}/c^2$~

\item $X(2120)$£º

\ \ \  ~$M=2122.4\pm6.7({\rm stat})^{+4.7}_{-2.7}({\rm syst})~{\rm
MeV}/c^2$~

\ \ \   ~$\Gamma=83\pm16({\rm stat})^{+31}_{-11}({\rm syst})~{\rm
MeV}/c^2$~

\item $X(2370)$£º

\ \ \  ~$M=2376.3\pm8.7({\rm stat})^{+3.2}_{-4.3}({\rm syst})~{\rm
MeV}/c^2$~

\ \ \  ~$\Gamma=83\pm17({\rm stat})^{+44}_{-6}({\rm syst})~{\rm
MeV}/c^2$~

\end{itemize}
For the $X(1835)$, the product branching fraction is
$B(J/\psi\rightarrow\gamma X(1835))\cdot
B(X(1835)\rightarrow\pi^{+}\pi^{-}\eta^\prime) = (2.87\pm0.09{\rm
(stat)}^{+0.49}_{-0.52}{\rm (syst)})\times10^{-4}$, and the angular
distribution of the  radiative photon is consistent with a
pseudoscalar assignment. The mass of the $X(1835)$ is consistent
with the BESII result, but the width is significantly larger. If we
fit the mass spectrum with one resonance as BESII, the mass and
width of the  X(1835) are $1841.2\pm2.9~{\rm MeV}/c^2$ and
$109\pm11~{\rm MeV}/c^2$, where the errors are statistical only.


In the mass spectrum fitting in Fig.~\ref{Fig2}(b), possible
interferences among different resonances and the non-resonant process are
not taken into account which might be a source of the large $\chi^2$
value for the fit ($\chi^2/d.o.f=144/62$). The dips around $2.2{~\rm
GeV}/c^2$ and $2.5{~\rm GeV}/c^2$ may not be fitted well because of
the neglect of such interferences. In the absence of knowledge of the
spin-parities of the resonances and their decay intermediate states,
reliable fits that include interference cannot be done.
  To determine the spin and parity of the
$X(1835)$, $X(2120)$ and $X(2370)$, and to measure their
masses and widths more precisely, a partial wave analysis must be
performed, which will be possible with the much higher statistics
$J/\psi$ data samples planned for future runs of the BESIII
experiment.

\vs 5mm

The BESIII collaboration thanks the staff of BEPCII and the
computing center for their hard efforts. This work is supported in
part by the Ministry of Science and Technology of China under
Contract No. 2009CB825200; National Natural Science Foundation of
China (NSFC) under Contracts Nos. 10625524, 10821063, 10825524,
10835001, 10935007; the Chinese Academy of Sciences (CAS)
Large-Scale Scientific Facility Program; CAS under Contracts Nos.
KJCX2-YW-N29, KJCX2-YW-N45; 100 Talents Program of CAS; Istituto
Nazionale di Fisica Nucleare, Italy; Russian Foundation for Basic
Research under Contracts Nos. 08-02-92221, 08-02-92200-NSFC-a;
Siberian Branch of Russian Academy of Science, joint project No 32
with CAS; U. S. Department of Energy under Contracts Nos.
DE-FG02-04ER41291, DE-FG02-91ER40682, DE-FG02-94ER40823; University
of Groningen (RuG) and the Helmholtzzentrum fuer
Schwerionenforschung GmbH (GSI), Darmstadt; WCU Program of National
Research Foundation of Korea under Contract No.
R32-2008-000-10155-0.

\ed